\documentstyle[preprint,aps,epsfig]{revtex}
\newif\iftightenlines\tightenlinesfalse
\tightenlines\tightenlinestrue

\def \be{\begin{equation}}
\def \ee{\end{equation}}
\def \bea{\begin{eqnarray}}
\def \eea{\end{eqnarray}}
\def \bit{\begin{itemize}}
\def \eit{\end{itemize}}
\def \Oi{{\mathcal O}}
\def \nnu{\nonumber}

\def \bqmm{b \to s\mu^+\mu^-}
\def \bqee{b \to s e^+e^-}
\def \bqtt{b \to s \tau^+\tau^-}
\def \bqll{b \to s l^+ l^-}

\def \fig#1{Fig.~\ref{#1}}
\def \br{{\cal B}\,}

\def \sec#1{Sec.~\ref{#1}}

\def \fig#1{Fig.~\ref{#1}}

\def \sec#1{Sec.~\ref{#1}}

\def\euro#1#2#3{{Eur. Phys. J. C} {\bf #1}, #3 (#2)}

\def\np#1#2#3{{Nucl.~Phys.}~{\bf B#1}, #3 (#2)}
\def\pl#1#2#3{{Phys.~Lett. B}~{\bf #1}, #3 (#2)}
\def\prd#1#2#3{{Phys.~Rev. D}~{\bf #1}, #3 (#2)}
\def\prl#1#2#3{{Phys.~Rev.~Lett.}~{\bf #1}, #3 (#2)}

\def\ptp#1#2#3{{Prog. Theor.~Phys.}~{\bf #1}, #3 (#2)}
\def\rmp#1#2#3{{Rev. Mod. Phys.} {\bf #1}, #3 (#2)}

\begin{document}

\preprint{\vbox{\baselineskip=14pt%
  \rightline{AMES-HET 03-01}
}}

\title{Rate difference between $\bqmm$ and $\bqee$ in SUSY with large
tan$\beta$}

\author{Yili Wang and David Atwood}

\address{{Department of Physics and Astronomy, Iowa State University,
Ames, IA 50011}}

\date{\today}
\maketitle
\begin{abstract}
We study the inclusive semileptonic rare decay $\bqll$ in minimal
supergravity model (mSUGRA). If tan$\beta$ is large, down-type quark mass
matrices and their Yukawa couplings cannot be diagonalized at the same
basis. This induces the flavor violating neutral Higgs boson couplings.
These couplings contribute significantly to decay $\bqmm$ and $\bqtt$, but
negligible to $\bqee$ decay because of its negligible $m_e$ mass. The
ratio $R \equiv B(\bqmm)/ B(\bqee)$ can be very different from its
corresponding value in the Standard Model. We find that part of parameter
space can accommodate a large R value, and that maximum R value can be
larger than 2. We also present our results in $\bqtt$ decay channel.
Although it can not be detected now, it is potentially a new channel for
the future observation of new physics.

\end{abstract}


\newpage

\section{Introduction}

Rare semileptonic B decays provide an extremely helpful tool to search new
physics beyond the Standard Model (SM). New physics contributions, which
enter through one loop radiative corrections, may be observed whenever the
SM contributions absent or suppressed. Thus, the measurement of $B
\rightarrow X_s l^+ l^-$ has a very good chance to reveal new physics
beyond Standard Model.

One way in which new physics may reveal its presence is for there to be a
deviation from the Standard Model prediction for $B \rightarrow X_s l^+
l^-$. In the absence of a deviation in the cross section, it is still
possible that new physics will be show itself in the details of the $B
\rightarrow X_s l^+ l^-$ signal. The detailed study of this signal can also
provide important clues as to the nature of the new physics.

In this paper, we suggest that a significant discrepancy between
$B\rightarrow X_s e^+ e^-$ and $B\rightarrow X_s \mu^+ \mu^-$ could arise
in some instances of the two Higgs doublet model. Indeed such a
discrepancy would strongly suggest new physics that couples to the mass of
the lepton. In this paper we discuss how observable signals of this sort
could arise in  the popular framework of minimal supergravity.

The
SM predicts the $\bqee$ and $\bqmm$ branching ratio for $M_{l^+l^-} > 2
m_\mu$ to be \cite{br1,br2}:

\be\label{sme}
 \br(\bqee) = 6.5 \times 10^{-6}
\ee
\be\label{smm}
 \br(\bqmm) = 6.2 \times 10^{-6}
\ee
\be\label{stt}
 \br(\bqtt) = 4.4 \times 10^{-7}
\ee
So that the ratio of branching fractions, $R_{SM}$, is:

\be\label{rsm}
 R_{\rm SM} \equiv \left. \frac{\br(\bqmm)}{\br(\bqee)} \right|_{\rm SM}
    \sim  .95
\ee

\noindent SM predicts $R_{SM} \sim$ 1 unless it is significantly altered
by new physics. Thus the measurement of $R$ provides a signal for specific
classes of new physics.

\begin{figure}[thb]
\begin{center}
\mbox{\epsfxsize=16cm\epsffile{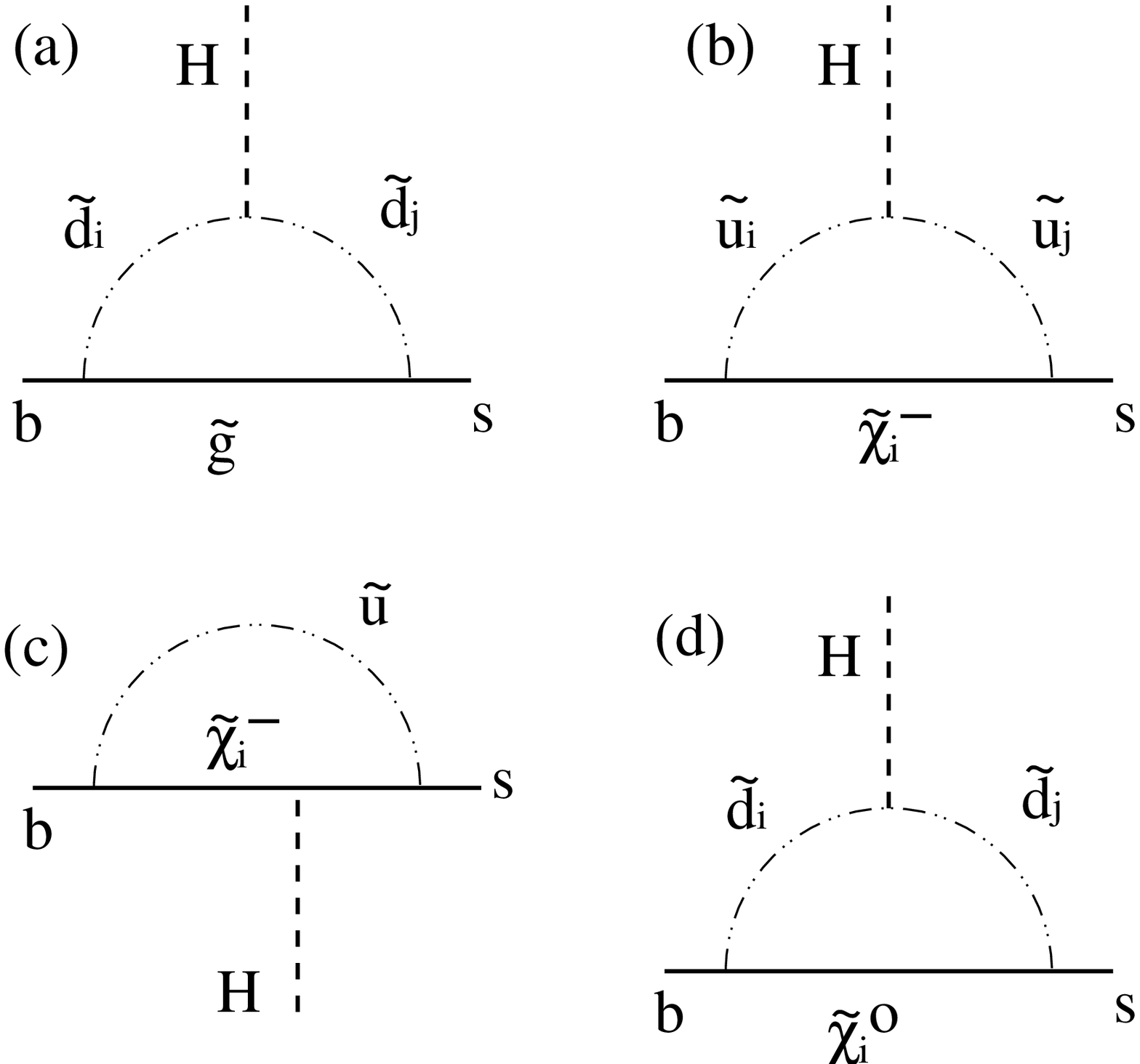}}
\vspace{3mm}
\caption{ One loop SUSY contributions.}
\label{fey}
\end{center}
\end{figure}

Supersymmetry (SUSY) is a promising candidate for new physics beyond
Standard Model \cite{rev}. In SM, the decay $\bqll$ occurs through
electroweak penguins and box diagrams. SUSY introduces several additional
classes of contributions showing in \fig{fey}:

\begin{enumerate}
\renewcommand{\labelenumi}{\alph{enumi}}
\item {gluino, down-type squark loop,} 
\item {chargino, up-type squark loop,} 
\item {chargino, up-type squark loop, (Higgs field attaching to charginos.)} 
\item  {neutralino down-type squark loop.}
\end{enumerate}

There are many papers exploring the photon penguin, Z penguin, gluino
penguin and box diagram contributions in SUSY \cite{TG,TG1,YG}. In our
paper, we focus on the penguin contributions to $\bqll$ mediated by
neutral Higgs bosons. At the one-loop level, couplings of the
\hspace{0.01cm} ``up-type''\hspace{0.01cm} Higgs field $H_u$ to down-type
quarks\cite{hrs,kolda,xt1} are induced. This coupling gives a new
contribution to the down type fermion mass matrix, and induces flavor
violating couplings of neutral Higgs bosons. This induced flavor violating
coupling increases with tan$\beta$. If tan$\beta$ is sufficiently large,
this flavor violating coupling has significant contribution to the
branching fraction of $B_s \rightarrow \mu^+ \mu^-$
\cite{gaur,chank,old,beku,isidori,dedes,arnowitt,bobeth2,buras,baek,huang}
and $\bqll$ decays \cite{br1,TG1,YG,eff,bll,bll1,ym,ch}. In this paper, we
would like to emphasize the contributions mediated by neutral Higgs
exchange. As we shall see, the ratio $R$ is particularly sensitive to new
physics which contributes through this kind of mechanism, because other
contributions which couple to $e$ and $\mu$ equally contribute to both the
numerator and denominator of Eq.\ref{rsm},\hspace{0.1cm} therefore giving
little contributions to R.
 
Higgs mediated SUSY contributions to the amplitude increase with lepton
mass $m_l$.  Therefore such contributions to $\bqee$ are negligible while,
given a sufficient large tan$\beta$, the contribution to $\bqmm$ may be
comparable to the SM and so observable deviations of $R$ from $R_{SM}$ are
possible. We will explore these effects as a function of SUSY parameter
space which, in some places, are even larger than the SM contributions. In
such cases, the ratio $R \equiv \frac{\br(\bqmm)}{\br(\bqee)}$ alters
substantially providing a possible way to detect new physics beyond SM.

The main object of this paper is to examine the prospects for observing a
deviation in $R$ within the framework of minimal supergravity (mSUGRA)
\cite{sug}. In \sec{ham}, we review the effective Hamiltonian for quark
transition $\bqll$. A detailed formula to calculate the branching fraction
of $\bqll$ is presented. \sec{msugra} contains a brief description of the
mSUGRA model along with our main results. We analyze the ratio R in mSUGRA
frame work and also discuss the large contribution of such effects to the
decay $\bqtt$. We present our conclusions in the last section.

\section{Effective Hamiltonian}\label{ham}
%

The effective Hamiltonian for $b \rightarrow s$ is derived by integrating
out the heavy degrees of freedom at the electroweak scale or above. This
Hamiltonian \cite{weak2,beku} can be written as:

\be\label{effect}
H = - \frac{G_F}{\sqrt{2}} V_{ts}^{*}V_{tb}^{}
[\sum_{i=1}^{10} C_i(\mu){\Oi_i}(\mu) + C_{Q_1}Q_1 + C_{Q_2}Q_2]
\ee

\noindent As in Ref.\cite{xt1,beku}, the operators we choose are in
basis\footnote{Only $\Oi_{7, 9, 10}$ and $Q_{1,2}$ are obviously relevant
to $\bqll$ directly.}:

\bea\label{operatorbasis}
{\Oi_1}&=& (\bar{s}_{ \alpha} \gamma_\mu P_L c_{ \beta})
(\bar{c}_{ \beta} \gamma^\mu P_Lb_{ \alpha}),\nnu\\
{\Oi}_2 &=& (\bar{s}_{ \alpha} \gamma_\mu P_L c_{ \alpha})
(\bar{c}_{ \beta} \gamma^\mu P_L b_{ \beta}),\nnu\\
{\Oi}_3 &=& (\bar{s}_{ \alpha} \gamma_\mu P_L b_{ \alpha})\sum_{q=u,d,s,c,b}
(\bar{q}_{ \beta} \gamma^\mu P_L q_{ \beta}),\nnu\\
{\Oi}_4&=& (\bar{s}_{ \alpha} \gamma_\mu P_L b_{ \beta})
\sum_{q=u,d,s,c,b}(\bar{q}_{ \beta} \gamma^\mu P_L q_{ \alpha}),\nnu\\
{\Oi}_5&=& (\bar{s}_{ \alpha} \gamma_\mu P_L b_{ \alpha})
\sum_{q=u,d,s,c,b}(\bar{q}_{ \beta} \gamma^\mu P_R q_{ \beta}),\nnu\\
{\Oi}_6&=& (\bar{s}_{ \alpha} \gamma_\mu P_L b_{ \beta})
\sum_{q=u,d,s,c,b}(\bar{q}_{ \beta} \gamma^\mu P_R q_{ \alpha}),\nnu\\ 
{\Oi}_7&=&\frac{e}{4 \pi^2}m_b (\bar{s}_{\alpha} \sigma_{\mu \nu}
P_R b_{\alpha})F^{\mu \nu},\nnu\\
{\Oi}_8& =& \frac{g_s}{4 \pi^2}m_b (\bar{s}_{\alpha} 
T_{\alpha \beta}^a \sigma_{\mu \nu}P_R b_{\beta}) G^{a \mu \nu},\nnu\\
{\Oi}_9&=& \frac{e^2}{4 \pi^2} (\bar{s}_\alpha \gamma^{\mu} P_L b_\alpha)
(\bar{l} \gamma_{\mu} l),\nnu\\
{\Oi}_{10}&=&\frac{e^2}{4 \pi^2} (\bar{s}_\alpha \gamma^{\mu} P_L
b_\alpha) (\bar{l} \gamma_{\mu}\gamma_5 l),\nnu\\
{Q}_1&=&\frac{e^2}{4 \pi^2}  (\bar{s}_L b_R) 
(\bar{l}l),\nnu\\
Q_2&=&\frac{e^2}{4 \pi^2} (\bar{s}_L b_R)
(\bar{l} \gamma_5 l),
\eea

$\alpha$, $\beta$ being color indexes, $a$ labels  the SU(3) generators,
and $P_{L,R}= (1\mp \gamma_5)/2$. $C_i$ are Wilson coefficients.

${\Oi}_1$ and ${\Oi}_2$ are current-current operators,
${\Oi}_3$...${\Oi}_6$ are QCD penguin operators.  The contributions of
these four quark operators to $\bqll$ are proportional to the tree level
matrix elements of operators ${\Oi}_7$, ${\Oi}_8$ and ${\Oi}_9$ at one
loop level. The chromomagnetic dipole operator, ${\Oi}_8$, gives no
contributions to $\bqll$. As a result, the ${\Oi}_1$ ...${\Oi}_6$
contributions can be absorbed by appropriately modifying the Wilson
coefficients of operator ${\Oi}_7$ and ${\Oi}_9$, which originated in the
$Z^0$ and $\gamma$ penguin diagrams with external $l^+ l^-$ pairs. The
neutral Higgs couplings SUSY contributions are mainly through $Q_1$ and
$Q_2$. The effective Hamiltonian for $\bqll$ is thus:

\bea\label{effect1}
H = - \frac{G_F \alpha}{\sqrt{2}} V_{ts}^{*} V_{tb}^{}
[ -  2 i C_7^{eff} \bar{s} \sigma_{\mu\nu}(m_b P_R + m_s P_L) 
\frac{q^{\nu}}{q^2} b \bar{l}\gamma^\mu l \nonumber 
+ C_9^{eff} \bar{s}_L \gamma_\mu b_L \bar{l} \gamma^\mu l \nonumber \\
+ C_{10} \bar{s}_L \gamma_\mu b_L \bar{l} \gamma^\mu \gamma_5 l 
 + C_{Q_1} \bar{s}_L b_R \bar{l} l + C_{Q_2} \bar{s}_L b_R \bar{l} \gamma_5 l]
\eea
    
The Wilson coefficient $C_9^{eff}$ includes leading-order (LO) and
next-to-leading order (NLO) logarithms, while $C_7^{eff}$ and $C_{10}$
enter only at the NLO level. In addition to the SM contributions to these
coefficients, there are several classes of SUSY contributions. The photon
penguin diagrams contribute to $C_7^{eff}$. Three types of diagram: the
photon penguin diagram; the Z penguin diagram; and the box diagram,
contribute to $C_9^{eff}$. The $C_{10}$ is induced by the Z penguin
diagram and the box diagram.

In the calculation of $C_7^{eff}$, we find that $C_7^{eff}$ can be quite
different from its SM value. There is a one-loop diagram with internal
stop and chargino which gives a large contribution when tan$\beta$ is
large. This stop-Higgsino diagram is proportional to the product of the
top and bottom Yukawa coupling constant, $m_t m_b/(sin\beta cos\beta)$,
which grows with tan$\beta$.  There are no such terms in the calculation
of $C_9^{eff}$ and $C_{10}$. The corresponding stop-Higgsino diagram is
proportional to the square of the top Yukawa coupling constants,
$m_t^2/sin^2\beta$, which does not grow for large tan$\beta$. As pointed
out by Ref.\cite{TG1}, SUSY contributions to $C_9^{eff}(m_b)$ and
$C_{10}(m_b)$ are very small and alter these coefficients by $\leq$ 5$\%$
over the whole parameter space on $C_9^{eff}(m_b)$ and $C_{10}(m_b)$.
Because of this, SUSY contributions on $C_9^{eff}$ and $C_{10}$ can be
ignored.

We must consider the SUSY contribution to $C_7^{eff}$. We write, $$
C_7^{eff}(M_w) = C_7^{SM}(M_w) + C_7^{SUSY}(M_w)$$ where $C_7^{SM}(M_w)$
is given in Ref.\cite{weak2}, and $C_7^{SUSY}(M_w)$ is taken from Ref.
\cite{YG,FG} with mass insertion approximation. For complete calculation,
see Ref.\cite{TG1}. At the $m_b$ scale: $$ C_7^{eff} = \eta^{16/23}
C_7^{eff}(M_w) + \frac{8}{3} (\eta^{14/23} - \eta^{16/23}) C_8(M_w) +
(\sum_{i = 1}^{8} h_i \eta^{a_i}) C_2(M_w)$$ where $\eta =
\frac{\alpha_s(M_w)}{\alpha_s(m_b)}$, $C_8(M_w)$, $C_2(M_w)$, $h_i$ and
$a_i$ can be found in Ref.\cite{weak2}. The expressions of $C_9^{eff}$,
and $C_{10}$ can be found in Refs \cite{beku,weak2}. Within SM, they are:

\be
C_9^{eff} = C_9 + Y(\hat s),\quad C_9 = 4.138,\quad C_{10} = -4.221 \nonumber
\ee

where $\hat s = q^2/m_b^2$, and $q$ denotes the invariant momentum of the
lepton pair. The expression for function Y($\hat s$) , coming from the one
loop contributions of operators ${\Oi}_1$ - ${\Oi}_6$, can be found in Ref
\cite{beku}. The Wilson coefficients $C_{Q_1}$ and $C_{Q_2}$, can be found
in Ref. \cite{xt1}. In terms of the Wilson coefficients, the differential
decay rate is \cite{eff,form}:

\bea\label{effect2}
\frac{d\Gamma}{d\hat s} = \frac{G_F^2 \alpha^2 m_b^5 }{128\pi^5}
|V_{ts}|^2 |V_{tb}|^2 \sqrt{\lambda(1,\mu_s, \hat s) 
\lambda(1,\frac{\mu_l}{\hat s}, \frac{\mu_l}{\hat s})}\quad \biggl[
 \frac{1}{6} (|C_9^{eff}|^2 + |C_{10}|^2) \nonumber \\ 
\biggr(\hat s (1 + \mu_s - \hat s)
\lambda(1,\frac{\mu_l}{\hat s},\frac{\mu_l}{\hat s}) 
+ (1 + \hat s - \mu_s)(1 - \hat s - \mu_s)(1 + \frac{2\mu_l}{\hat s})\biggr) 
\nonumber \\
 + (|C_9|^2 - |C_{10}|^2)\mu_l (1 + \mu_s - \hat s) + 
\frac{2}{3}|C_7^{eff}|^2 \frac{1}{\hat s} (1 + \frac{2\mu_l}{\hat s})  
\nonumber \\
\biggr(2(1 + \mu_s)(1 - \mu_s)^2 - \hat s 
(1 + 14 \mu_s + \mu_s^2)-{\hat s}^2 (1+\mu_s)\biggr) \nonumber \\
+2 C_7^{eff} Re(C_9^{eff})(1 + \frac{2\mu_l}{\hat s})
\biggr((1-\mu_s)^2 -\hat s (1+ \mu_s)\biggr) \nonumber \\
+ \frac{1}{4}C_{Q_1}^2 (1 + \mu_s - \hat s)(\hat s - 4 \mu_l) 
+ \frac{1}{4}C_{Q_2}^2 {\hat s}(1 + \mu_s - \hat s) \nonumber \\
+ C_{10}C_{Q_2} \sqrt{\mu_l} (1 - \mu_s - \hat s )\biggr]
\eea 

In Eq.\ref{effect2}, $\mu_l = m_l^2 / m_b^2$, and $\mu_s = m_s^2 / m_b^2$. 
The function $\lambda(x, y, z)$ is:
\bea
\lambda(x, y, z) = x^2 + y^2 + z^2 - 2xy - 2xz - 2yz \nonumber 
\eea

Integrating out $\hat s$ from $4 m_l^2/m_b^2$ to $(1-m_s/m_b)^2$, 
the branching ratios of decay $\bqll$ are easily obtained by:
\bea\label{branching}
\br(\bqll) = \frac{1}{\Gamma} {\int_{4 m_l^2/m_b^2}^{(1 - m_s / m_b)^2} 
\left (\frac{d \Gamma}{d \hat s} \right ) d \hat s} 
\label{brac}
\eea
 
\section{Results}\label{msugra}
The minimal supergravity model \cite{sug} has been a popular model for
SUSY phenomenology. It provides a well motivated realization of the
Minimal Supersymmetric Standard Model (MSSM). In this model, SUSY is
broken in a hidden sector. The fields in the hidden sector interact with
usual particles and their superpartners only via gravity. In this way,
SUSY breaking is communicated to the observable sector of Standard Model
particles and their superpartners.

The mSUGRA framework assumes that at the GUT scale ($M_{GUT} \approx 2
\times 10^{16}~GeV$), all scalar fields have a common SUSY breaking mass
$m_0$ , all gauginos have a mass $m_{\frac{1}{2}}$, and also all soft SUSY
breaking trilinear scalar couplings have a common value $A_0$. The
resulting MSSM will have various soft SUSY breaking terms unified at
$M_{GUT}$. The soft SUSY breaking parameters are evolved from $M_{GUT}$ to
weak scale using renormalization group equations (RGE). Minimization of
the Higgs potential gives a relation between $\mu^2$, $m_Z^2$, B and
tan$\beta$, therefore $\mu^2$ can be determined by $m_Z^2$ \cite{mu} and
the B parameter is traded in favor of tan$\beta$, so that the model with
radiative electrical symmetry breaking is completely specified by the SM
parameters together with: $$m_0,\quad m_{\frac{1}{2}},\quad A_0,\quad
tan\beta,\quad sign(\mu)$$ To evaluate the mass spectrum of the MSSM
resulting from mSUGRA, we use the ISASUGRA program from the ISAJET
package\cite{isajet}. With these sparticle masses and mixing angles, we
can get the Wilson coefficients $C_{Q_1}$ and $C_{Q_2}$. Then the
branching ratio of $\bqll$ can be computed using Eq. \ref{effect2} and
\ref{branching}.
      
We calculated the branching ratios of $\bqee$ and $\bqmm$. We find that if
tan$\beta$ is small, SM contributions dominate and SUSY contributions are
very small. As tan$\beta$ grows, SUSY contributions become large. $O_7$
gives a sizable SUSY contributions with a little dependent on the lepton
mass.  The $O_7$ and $O_9$ interfere term also gives a comparable
contributions. They together alter the branching fraction around $15\%$ to
$20\%$. $Q_1$ and $Q_2$ terms are induced by Higgs-mediated contributions.
Their contributions to $\bqll$ branching fraction increase with
$tan^6\beta$ and in some parameter space, they are even bigger than SM
contributions. The corresponding Wilson coefficients $C_{Q1}$ and $C_{Q2}$
are proportional to the lepton mass so that the branching ratio is
proportional to lepton mass square. Therefore, Higgs-mediated SUSY
contributions in decay $\bqmm$ are about $10^4$ bigger than that in
$\bqee$ decay and alter the ratio $R \equiv B(\bqmm)/ B(\bqee)$ deviating
from its corresponding SM value. We require that the lepton pair mass is
larger than 2 $m_{\mu}$ to reduce the dependences of the branching
fraction on the lepton pair mass in small s range.

\begin{figure}[thb]
\begin{center}
\mbox{\epsfxsize=16cm\epsffile{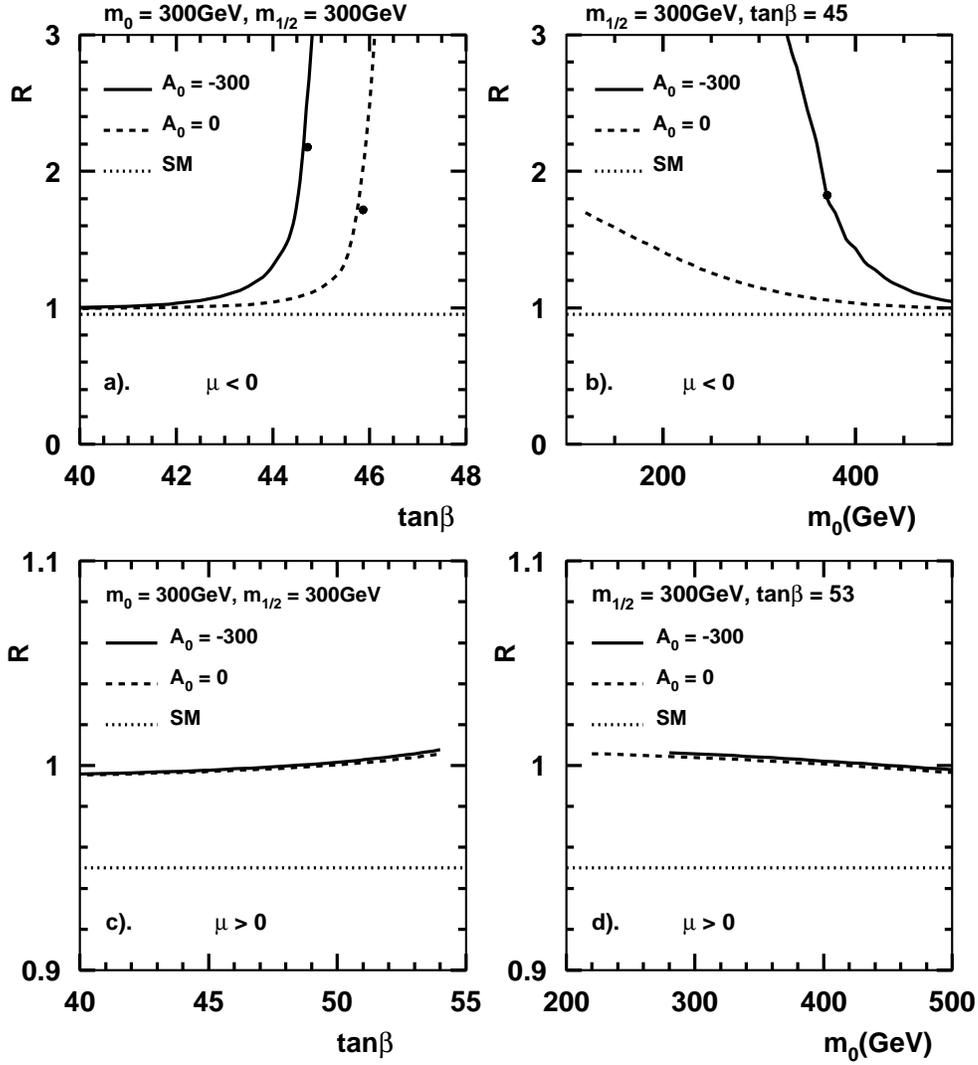}}\hspace{0.2cm}
\vspace{3mm}
\caption{Ratio $R \equiv B(\bqmm)/B(\bqee)$ in mSUGRA model with $ m_{1/2} = 300$~GeV and a) R vs $tan\beta$ with $\mu < 0, m_0 = 300$~GeV, b) R vs $m_0$ with $ \mu < 0, tan\beta = 45$, c) R vs $tan\beta$ with $\mu > 0, m_0 = 300$~GeV, d) R vs $m_0$ with $ \mu > 0, tan\beta = 53$. The solid line is for $A_0 = -300$~GeV, and $A_0 = 0$~GeV for dashed line. The dotted line is the corresponding SM line. The circles mark the limits of the experimentally allowed regions discussed in the text.}
\label{tanb}
\end{center}
\end{figure}

In \fig{tanb}, we show the dependence of $R \equiv B(\bqmm)/B(\bqee)$ on
tan$\beta$ and $m_0$ for $A_0 = 0$ and $A_0 = -300$ cases. We fixed $
m_{1/2} = 300$~GeV in all the frames. Frame a). is R vs. tan$\beta$ with
$\mu < 0$ and $m_0=300$~GeV. Frame b) is R vs. $m_0$ with $\mu < 0$ and
tan$\beta$ = 45. Frames c) and d) are for $\mu>0$ case. Frame c) is R vs.
tan$\beta$ with $m_0 = 300$~GeV, while frame d) has tan$\beta = 53 $ with
R vs $m_0$ plot. The solid (dashed) line is for $A_0 = -300$ (0) GeV, and
the dotted line is the corresponding R value in SM. Values of tan$\beta$
($m_0$) larger than the corresponding values denoted by circles on the
curves are where $m_h$ falls below its experimental bound 113 GeV, which
gives out the most stringent constraint. In frame b), $m_h > 113$~GeV for
all the regions in the graph along $A_0 = 0$ line. In frames c) and frame
d), $m_h$ is always larger than 113 GeV.  The main reason to show this
figure is to understand the behavior of the ratio R.  Several features
need to be noted:

\begin{itemize}

\item The $R$ value is significantly larger for negative value of $\mu$.
As explained in ref. \cite{xt1}, changing the sign of $\mu$ changes the
denominator of $\chi_{FC}$ in Eq.10 of ref. \cite{xt1}, so that a
suppression for positive $\mu$ changes to enhancement for negative $\mu$.

\item For $\mu < 0$ case, in low tan$\beta$ regions, SM contributions
dominate, so that all curves are close the SM value: R $\sim$~0.95. Same
thing happens in $m_0$ larger than 400 GeV range.

\item Again in $\mu < 0$ case, as tan$\beta$ is larger than 44, there is a
sharp rise for both $A_0 = 0$ and $A_0 = -300$. This is because
Higgs-mediated SUSY contributions are expected to behave as $tan^6\beta
/m_A^4$\cite{xt1} \footnote{The contribution from h exchange is very small
as long as $h$ is a SM-like Higgs boson, and $m_H \sim m_A$.}.  The solid
and dashed curves indeed show this behavior as long as tan$\beta$ is
large. Clearly then, SUSY contributions are dominated by Higgs mediated
penguin when tan$\beta$ is large.

\item R deviates significantly from the SM value when tan$\beta$ larger
than 45 in $\mu < 0$ case. Unfortunately, in the mSUGRA scenario, these
ranges are excluded by the experiment bound on the Higgs mass, $m_h >
113$~GeV \cite{lep}.

\item As shown in frame c) and frame d), in $\mu > 0$ case, R changes very
slowly over the whole parameter space. Higgs-mediated SUSY contributions
are very small \cite{xt1}. The fact that solid line and dashed line are
almost overlap and parallel to the dotted line tells us a constant R value
exists. We even increase the value of $A_0$ to around 300~GeV. Still we get
a $R$ value, which is very close to its SM prediction. This excludes the
possibility to find a signal of new physics in $\mu > 0$ case.

\end{itemize}

Fig. \ref{tanb} illustrates that typically large contributions to $R$
occur in the $\mu < 0$ scenario with tan$\beta \geq 40$.  For $\mu > 0 $,
R is almost constant over the whole parameter space and very close to the
SM prediction, so that no room has left for new physics. Although $\mu <
0$ is generally thought to be disfavored by the determination of the muon
anomalous magnetic moment by the E821 experiment \cite{e821}, a
conservative estimate \cite{km} of the theoretical error suggests that
there is a region allowed \cite{baer} by this constraint, though perhaps
in conflict with $B(b \to s\gamma)$ \footnote{It is worth reminding the
reader that SUSY contribution, special for large values of tan$\beta$, may
have considerable theoretical uncertainty. Unlike constraints from direct
searches, constraints from $B(b \to s \gamma)$ are very sensitive to
details of the model. A small amount of flavor mixing in the squark
sector could lead to large differences in the predictions of $b \to s
\gamma$ decays.  For this reason, the constraints from $B(b \to s\gamma)$
should be interpreted with some care.}, where $b \to s l^+l^-$ may provide
the first hint of new physics if tan$\beta \geq 40$. This region would
expand as more experimental data are accumulated.

\begin{figure}[thb]
\begin{center}
\mbox{\epsfxsize=16cm\epsffile{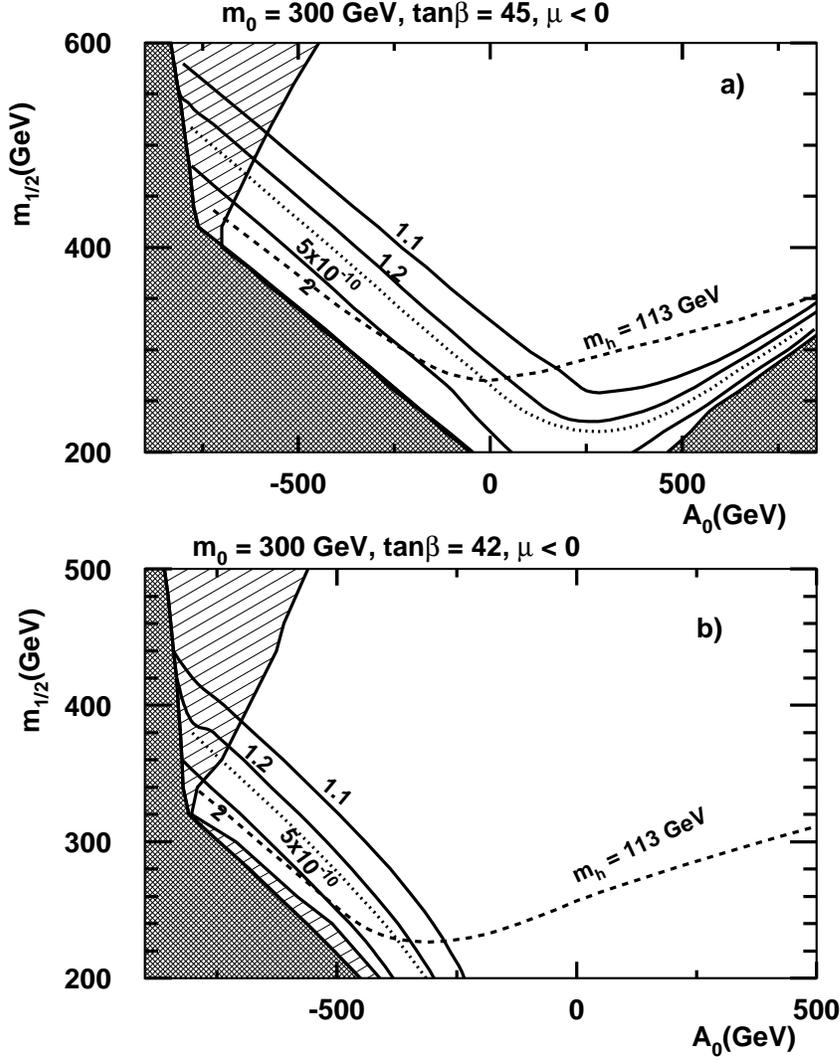}}\hspace{0.2cm}
\vspace{3mm}
\caption{Contours of constant ratio $R \equiv B(\bqmm)/B(\bqee)$ with $\mu < 0, m_0 = 300$~GeV in $A_0 - m_{1/2}$ plane with a) $tan\beta = 45$ b) $tan\beta = 42$ in mSUGRA model. The dark-shaded region is excluded by the theoretical constraints on electroweak symmetry breaking (EWSB). and the slant-hatched region is excluded for $\tilde Z_1$ is not LSP. Below the dashed line is for $m_h < 113$~GeV. Below the dotted line, the branching fractions of  $B_s \to \mu^+ \mu^-$ decay are larger than $5 \times 10^{-10}$. }
\label{con}
\end{center}
\end{figure}

The large values of $R$ in frame a) and b) in Fig. \ref{tanb} encourage us
to do further exploration in $\mu < 0$ case. Therefore, we plot a contours
of R values in $A_0$ - $m_{1/2}$ plane in \fig{con} for $\mu < 0$ case. We
choose $m_0 = 300$~GeV and in frame a), tan$\beta = 45$, while in frame
b), tan$\beta = 42$.  The dark-shaded regions are excluded on theoretical
grounds because the overall theory does not lead to electroweak symmetry
breaking. The slant-hatched region are excluded because of $\tilde Z_1$ is
not the lightest supersymmetric particle (LSP). If the neutralino is the
LSP, it will be stable and therefore a dark matter candidate. Within the
slant-hatched region, charged sparticle is the LSP which disagrees with
cosmologies models.  Below the dashed line labeled \hspace{0.1cm} ``$m_h =
113$~GeV''\hspace{0.1cm} are the regions $m_h$ is smaller than its
experimental bound 113 GeV. Below the dotted line, branching fractions of
$B_s \to \mu^+ \mu^-$ are larger than $5 \times 10^{-10}$, but smaller
than $10^{-7}$. The later is the maximum limit the experiment, Tevatron or
B factories, can explore in detecting $B_s \to \mu^+ \mu^-$ decay. The
contours of $R \equiv B(\bqmm)/B(\bqee)$ are labeled by the values of
corresponding ratio R. From frame a) we see that in the allowed region, R
can be larger than 2. Even in frame b), R can be as large as 2.  The
outermost curves is for R equals to 1.1. Turning our attention to the
sensitivity with respect to tan$\beta$, we show in frame b), the same
contours, but with tan$\beta = 42$. Although the range for tan$\beta = 42$
is smaller than that for tan$\beta = 45$, it has still regions where B
factories may be expected to detect it. The ratio can be significantly
larger below the R = 2 contour.

Recently, Belle experiment has observed a signal for exclusive B decays
with branching fraction $\br(B \rightarrow K l^+ l^-) =
(0.75_{-0.21}^{+0.25} \pm 0.09) \times 10^{-6}$ averaged over electron and
muon channels and $\br(B \rightarrow K \mu^+ \mu^-) =
(0.99_{-0.32-0.14}^{+0.40+0.13}) \times 10^{-6}$ in muon channel
\cite{belle}. This is consistent with our SM value for the inclusive rate
$\sim 6.4 \times 10^{-6}$. The inclusive B decay branching ratio is
measured by Belle group with $\br(B \to X_S l^+ l^-) = (6.1 \pm
1.4(stat)^{+1.4}_{-1.1}(syst)) \times 10^{-6}$ \cite{bell2,bell1}, which is
consistent with the SM prediction. We also estimate that approximately
$\sim 10^{7}$ B mesons are required to reach the $R = 1.2$ limit.

\begin{figure}[thb]
\begin{center}
\mbox{\epsfxsize=16cm\epsffile{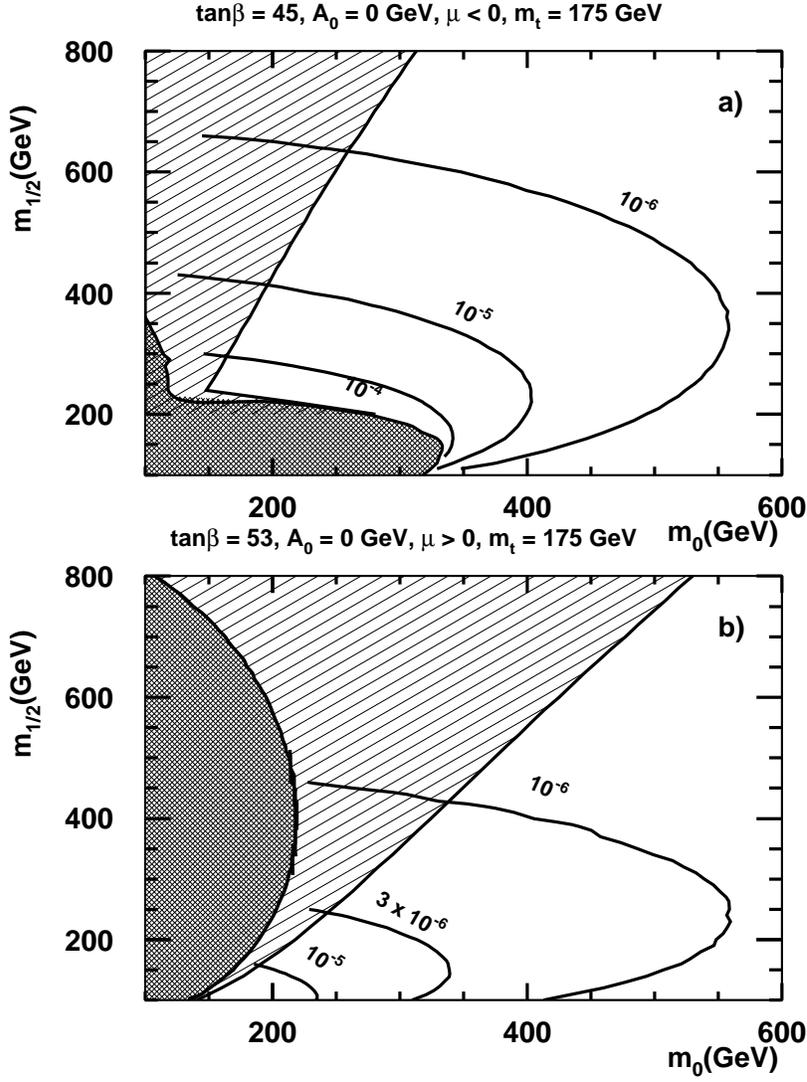}}
\vspace{3mm}
\caption{Contours of constant branching fraction for decay $\bqtt$. The results are showed in $m_0 - m_{1/2}$ plane with $A_0 = 0$~GeV, in frame a). $tan\beta = 45$, $\mu < 0$ and in frame b). $tan\beta = 53$,  $\mu > 0$. The dark-shaded region is excluded by the theoretical constraints on EWSB. and the slant-hatched region is excluded for $\tilde Z_1$ is not LSP.}
\label{tau}
\end{center}
\end{figure}

Due to the $m_l^2$ dependence of these branching fractions, one would
clearly expect contributions to $\bqtt$ \cite{form,tau1} two orders of
magnitude greater than $\bqmm$. Turning to \fig{tau}, we see that this is
indeed the case. In \fig{tau}, we show the contours of $B(\bqtt)$ in $m_0
- m_{1/2}$ plane with $A_0 = 0$~GeV. In frame a) $\mu < 0$ and tan$\beta =
45$, while in frame b) $\mu > 0$ and tan$\beta = 53$. As in \fig{con}, the
dark-shaded regions are excluded by theoretical constraints:
charge-breaking minimal or lack of electroweak symmetry breaking. $\tilde
Z_1$ is not the lightest supersymmetric particle (LSP) in the
slant-hatched region. The contours are labeled by the corresponding values
of $\bqtt$ branching fraction. The outermost contour corresponds to a
branching fraction of $10^{-6}$.  The SM branching fraction is around
$B(\bqtt) \sim 3.66 \times 10^{-7}$. In \fig{tau}, at the most parameter
space, SUSY contributions is much larger than the SM background, reaching
a limit of $\sim 10^{-4}$ within the allowed region. Here we remind the
readers that some region perhaps is excluded by $B_S \to \mu^+ \mu^-$
constraint. In ref.\cite{xt1}, we give out detailed calculations on $B_s
\to \mu^+\mu^-$.  Unfortunately $\tau$ identification is very difficult.
The branching fraction about $10^{-3}$ for $\tau$ process may be accessed
at B factory. Clearly then, at least one order of magnitude improvement in
the acceptance for $\bqtt$ is required to have a significant impact on
this parameter space. It is interesting to note that any new physics which
contribute significantly to $\bqtt$ will also give a signal in $B_s
\rightarrow \tau\tau$ as discussed in Ref.\cite{xt1}.

Another aspect of new physics contributions to the $\bqmm$ signal is the
forward-backward lepton asymmetry $A_{FB}$. This is considered in ref.
\cite{ch} and found to be on the order of a few percent hence difficult to
observe at current B factory luminosities. We do not explore this feature
in this paper, but it is potentially another channel for new physics
detection.

\section{Summary and Conclusions}\label{discussion}

We have explored the rare semileptonic decay $\bqee$, $\bqmm$ and $\bqtt$
within mSUGRA framework where some Higgs-mediated SUSY contributions to
the branching fraction of $b \to l^+l^-$ are proportional to corresponding
lepton mass squared.  When such contributions in the muon case become
comparable to the SM, this opens a new channel for new physics detection.
We explored these decays in detail and find that while in $\mu > 0$ case,
SUSY contribution is suppressed, for negative $\mu$ with large tan$\beta$
the flavor violating couplings of neutral Higgs bosons to down type quarks
leads to substantial SUSY contributions to the branching fraction of decay
$\bqmm$.  In particular observably large contributions to the branching
fraction of $\bqmm$ are possible when tan$\beta > 40$. In particular there
can be leading a large discrepancy between the $\bqmm$ and $\bqee$
branching ratios.  The resultant deviations in $R \equiv
B(\bqmm)/B(\bqee)$ from the SM values in some regions of the mSUGRA
parameter space are as shown in \fig{con}.

In this framework the large value of $m_\tau$ would give a huge deviation
from the SM prediction for $B(\bqtt)$ in the region of large tan$\beta$.  
We explore the branching fraction for $\bqtt$ decay in $m_0 - m_{1/2}$
plane and in the region of parameter space considered the maximum values
of $B(\bqtt)$ can be up to $8 \times 10^{-4}$. Unfortunately, because of
the difficulties in separating signals of $\tau$ decay from the
backgrounds, the $\tau$ signal is undetectable at B factories. Thus,
the most promising signal of the new physics considered in this paper is a deviation in the value of $R$ from its SM value.
Conversely, if such a deviation in $R$ were observed it would strongly
indicate new physics coupling to mass through the Higgs sector as in the
case of mSUGRA model considered in this paper.

\section*{Acknowledgments}

It is a pleasure to thank Professor X. Tata for his helpful comments. We
would also like to thank Professor G. Hillar for several useful
discussions on the subject of this paper.  Thanks Professor G. Valencia
for the discussion on this paper and also thank Professors J. Cochran and
S. Prell for valuable information about the reach of BaBar experiments to
$\bqll$ decays, and for providing us with Ref.\cite{barbar}. This work is
supported by DOE under contact number DE-FG02-01ER41155.



%
\iftightenlines\else\newpage\fi
\iftightenlines\global\firstfigfalse\fi
\def\dofig#1#2{\epsfxsize=#1\centerline{\epsfbox{#2}}}
\def\fig#1#2{\epsfxsize=#1{\epsfbox{#2}}}

\newpage
\newpage

\newpage

\newpage

\newpage
\end{document}